\documentclass[nohyper,letterpaper,11pt,notoc]{JHEP3}
\usepackage{graphicx}
\usepackage{amssymb}
\usepackage{amsmath}
\usepackage{amsfonts}
\usepackage{latexsym}
\usepackage[all]{xy}
\usepackage{slashed}
\usepackage{slashbox}

\preprint{}

\newcommand{\MeV}{~\mathrm{MeV}}
\newcommand{\GeV}{~\mathrm{GeV}}
\newcommand{\TeV}{~\mathrm{TeV}}

\newcommand{\mgam}{m_b}

\newcommand{\uoned}{U(1)_d}
\newcommand{\gd}{g_d}
\newcommand{\cO}{\mathcal{O}}
\newcommand{\be}{\begin{eqnarray}}
\newcommand{\ee}{\end{eqnarray}}

\title{Lepton Jets in (Supersymmetric) Electroweak Processes}
\author{Clifford Cheung$^{(a)}$, Joshua
  T. Ruderman$^{(b)}$, Lian-Tao Wang$^{(b)}$ and Itay
  Yavin$^{(b,c)}$\\ \it{(a) School of Natural Sciences, Institute for
    Advanced Study, Princeton, NJ 08540}  \\ \it{(b)
    Department of Physics, Princeton University, Princeton, NJ 08544} \\ \it{(c)
    Center for Cosmology and Particle Physics, Department of Physics, New York University, New York, NY 10003}} 

\abstract{We consider some of the recent proposals in which weak-scale dark matter is accompanied by a GeV scale dark sector that could produce spectacular lepton-rich events at the LHC.  Since much of the collider phenomenology is only weakly model dependent it is possible to arrive at generic predictions for the discovery potential of future experimental searches.  We concentrate on the production of dark states through $Z^0$ bosons and electroweak-inos at the Tevatron or LHC, which are the cleanest channels for probing the dark sector.  We properly take into account the effects of dark radiation and dark cascades on the formation of lepton jets. Finally, we present a concrete definition of a lepton jet and suggest several approaches for inclusive experimental searches.
}

\begin{document}
\section{Introduction} 

In light of recent astrophysical observations, the authors of \cite{ArkaniHamed:2008qn} have proposed a broad class of theories in which the annihilation of $\sim$TeV scale dark matter in the galactic halo accounts for the anomalous excess of cosmic ray leptons.  While this dark matter is probably inaccessible to colliders, it is accompanied by a $\sim$GeV scale dark sector which couples, albeit very weakly, to the standard model (SM).  The dark sector states are relatively light and can be produced at high energy colliders, only to cascade decay through the dark sector and ultimately return to the visible sector as electrons, muons, and possibly pions.  The outgoing leptons emerge in the detector as lepton jets~\cite{ArkaniHamed:2008qp}, which are highly collimated multiple muons or electrons that result from the decay of these highly boosted dark sector states.  As observed in ref.~\cite{ArkaniHamed:2008qp} and later elaborated on in ref.~\cite{Baumgart:2009tn}, the addition of supersymmetry also bears with it a number of interesting and novel collider signatures. Some of the phenomenology is also relevant for scenarios with hidden valleys~\cite{Strassler:2006im} or where DM decays into a $\GeV$ scale sector \cite{Ruderman:2009tj}, and as such may describe a large class of models. Recently, experimental effort in searching for such objects has been reported in \cite{Abazov:2009yi}. 

\begin{figure}[h]
\begin{center}
\includegraphics[width=0.6 \textwidth,height=0.3\textheight]{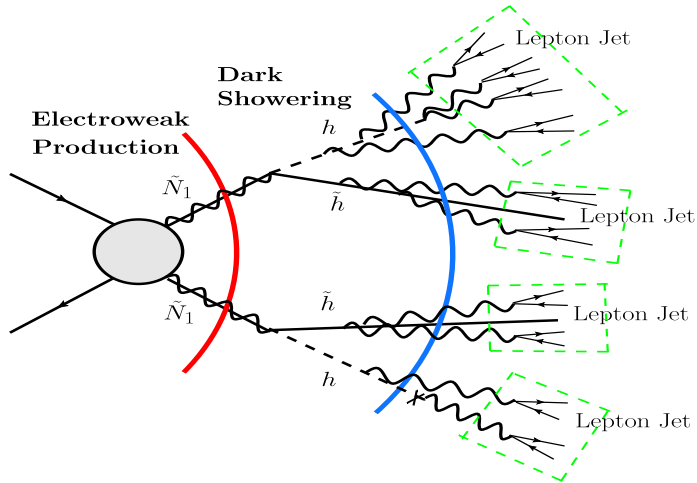}
\end{center}
\caption{A schematic illustration of the type of events we consider in this work. The time evolution can be divided into three stages: electroweak boson or -ino production and subsequent decay into the dark-sector, evolution through the dark sector, and finally the formation of lepton jets, as delineated by the dashed boxes. Such events may also include missing energy.}
\label{fig:threestage}
\end{figure}

The aim of the present work is to provide a quantitative study of the collider phenomenology of a typical event involving the GeV scale sector and resulting in lepton jets. We consider the minimal scenario whereby the SM sector and the dark sector are coupled only via kinetic mixing terms as detailed below. Effectively, dark sector fermions and scalars have a small coupling to the $Z^0$ boson and the minimal supersymmetric standard model (MSSM) bino, while the dark sector gauge boson couples weakly to the SM electromagnetic current.  As a result of these couplings, a typical event factorizes into three modular stages, which are illustrated in Fig.~\ref{fig:threestage}:
\begin{itemize}
\item {\bf Electroweak Production}

Once $Z^0$ bosons and electroweak-inos are produced, they decay into the dark sector via the aforementioned coupling. Rare $Z^0$ decays and electroweak-ino pair production are the cleanest channels in which to observe the production of dark sector states due to the limited hadronic activity. Decays into the dark sector depend weakly on the dark sector details.

\item {\bf Dark Sector Evolution}

Once the electroweak states have decayed into the dark sector, the resulting dark states are highly boosted and cascade decay down to the bottom of the dark sector spectrum.  A universal feature of these dark matter scenarios is that there is log enhanced soft emission of dark gauge bosons, which increases the multiplicity of light dark sector states and ultimately yields a greater number of final state leptons.  We simulate these dark sector showers and discuss their characteristics.   

\newpage
\item {\bf Outgoing Lepton Jets}

Cascade decays and soft emissions through the dark sector result in radiated dark gauge bosons which return to the visible sector as collimated lepton jets. After studying the shape and distribution of simulated lepton jets and taking into account the possible dilution and contamination from the decay of the dark bosons into pions,  we suggest a concrete definition for a lepton jet which can be used in inclusive experimental searches for these objects. While much of the phenomenology we consider is independent of the details and spectra of the particular dark sector model, this is not the case for the bottom of the dark sector spectrum, which may be probed by studying lepton jet shapes.

\end{itemize}

In section \ref{sec:ewprod} we review how the dark sector couples to the visible sector and discuss the production of dark sector states in rare $Z^0$ decays at LEP, the Tevatron and the LHC.  We also consider electroweak-ino pair production at the Tevatron and LHC. In section \ref{sec:dynamics} we consider the evolution in the dark sector which includes dark showering and cascade decays in the dark sector itself. Section \ref{sec:lepjet} begins with an analysis of the final state leptons and the formation of lepton jets and ends with some proposals for experimental searches. Section \ref{sec:conclusions} contains our conclusions.  

\section{Electroweak Production}
\label{sec:ewprod}

Let us first review how the visible sector and dark sector are coupled.  For a detailed treatment, see \cite{Baumgart:2009tn}.  As in \cite{ArkaniHamed:2008qn}, we assume the existence of a new dark gauge group which contains a $U(1)$ factor that is spontaneously broken at $\sim$GeV.  The associated dark gauge boson kinetically mixes with the SM photon and the $Z^0$ vector-boson.  We parameterize these couplings as
\be
\label{eqn:kinmix}
{\cal L}_\textrm{gauge mix} &=& -\frac{1}{2}\epsilon_1 b_{\mu\nu} A^{\mu\nu} - \frac{1}{2} \epsilon_2  b_{\mu\nu} Z^{\mu\nu} \\ &=&-\frac{1}{2}\epsilon_1^\prime b_{\mu\nu} B^{\mu\nu} - \frac{1}{2} \epsilon_2^\prime  b_{\mu\nu} W_3^{\mu\nu}
\ee
where $b_{\mu\nu}$ denotes the field strength for the dark gauge boson and $\epsilon_{1,2}$ and $\epsilon_{1,2}^\prime$ are related by the Weinberg angle. In particular, when only $\epsilon_1^\prime$ is present, we have $\epsilon_1 = \epsilon_1^\prime\cos\theta_W$ and $\epsilon_2 = \epsilon_1^\prime\sin\theta_W$. The first parameterization (mass basis) is more useful when discussing SM processes while the second (gauge basis) comes in handy when considering the MSSM\footnote{The operator involving $W_3$ is not gauge-invariant with respect to $SU_L(2)$. It should be thought of as the result of a higher dimensional operator $b_{\mu\nu} tr(H^\dag W^{\mu\nu} H)/\Lambda^2$, where $\Lambda$ is some high-scale. After the higgs condenses one obtains an effective mixing between $b_{\mu\nu}$ and $W_{3 \mu\nu}$.}.  With the addition of supersymmetry, the above coupling implies a kinetic mixing between the dark bino, ${\tilde b}$ and the MSSM gauginos, ${\tilde B}$ and ${\tilde W_3}$:
\be
{\cal L}_{\textrm{gaugino mix}} &=& -2i\epsilon_1' {\tilde b}^\dagger \bar\sigma^\mu \partial_\mu {\tilde B}   -2i\epsilon_2' {\tilde b}^\dagger \bar\sigma^\mu \partial_\mu {\tilde W_3}+{\rm h.c.}
\ee
The gauge and gaugino kinetic mixings can both be eliminated by a set of field redefinitions which induce the portal to the dark sector which will be relevant to this collider study:
\be
\label{eq:portal}
{\cal L}_{\rm portal} &=& \epsilon_1 b_\mu J_{\rm EM}^\mu + \epsilon_2 Z_\mu J_b^\mu + \epsilon_1' {\tilde B} \tilde J_{\tilde b}+\epsilon_2' {\tilde W}_3 \tilde J_{\tilde b} \\
 J_b^\mu &=& g_d \sum_i q_i\left(i (h_i^\dagger \partial^\mu h_i -h_i \partial^\mu h_i^\dagger ) + \tilde h_i^\dagger \bar{\sigma}^\mu \tilde h_i\right)
\\
\tilde{J}_{\tilde b} &=& -i\sqrt{2} g_d \sum_i q_i \tilde{h}_i^\dagger h_i
\ee
where $J_{\rm EM}$ is the SM electromagnetic current and $J_b$ and $\tilde J_{\tilde b}$ are the bosonic and fermionic components of the dark gauge current.  In particular, $J_b$ contains dark scalar and dark fermion bilinears, while $\tilde J_{\tilde b}$ contains mixed dark scalar-fermion bilinears. 

Since the dark sector scalars and fermions couple to the $Z^0$ boson and the MSSM bino, they will be produced in rare $Z^0$ decays and electroweak-ino decays, which we now consider in detail. 

\subsection{Rare  $Z^0$ Decays}
\label{sec:zdecay}

The strongest bound on kinetic mixing with a new $\sim$GeV scale vector-boson comes from the muon $g-2$ ratio~\cite{Pospelov:2008zw}, which constrains the photon mixing to be 
\begin{equation}
\epsilon_1^2 \lesssim 3 \times 10^{-5}\frac{m_b}{100\MeV}
\end{equation}
On the other hand, the branching ratio of $Z^0$ into a pair of dark fermions charged under the dark gauge group is fixed by the mixing parameter, $\epsilon_2$, which is not directly limited by this bound and can be somewhat larger. The branching fraction of $Z^0$ to dark fermions is 
\begin{equation}
\label{eqn:ZBR}
{\rm BR}\left(Z\rightarrow f\bar{f}\right) = \frac{\epsilon_2^2 g_d^2}{12\pi} \frac{M_{Z^0}}{\Gamma_{Z^0}}
\end{equation}
while the branching fraction into a pair of dark scalars is a quarter of this.  This has implications for LEP, as well as the Tevatron and LHC:

{\bf LEP:} With a nominal value of $\alpha_d = \alpha_{\rm EM}(M_{Z})$ and $\epsilon_2^2 = 10^{-4}$ the 17M on-shell $Z^0$ bosons produced at LEP~\cite{Z-Pole} will yield about 150 events of dark fermion pairs. While the LEP data can at the very least place stronger bounds on the couplings, it is not clear that existing searches covered the type of event topologies involved. Collimated leptons coming off of dark sector states may have been missed in exclusive searches requiring isolated leptons. 
 
\begin{figure}[h]
\begin{center}
\includegraphics[scale=.5]{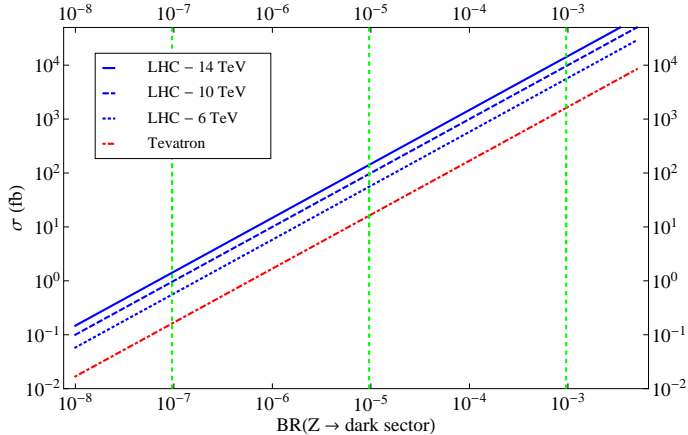}
\end{center}
\caption{The cross-section for production of dark sector states via $Z^0$ as a function of the branching ratio for both the Tevatron and LHC. The vertical green dashed lines mark the branching ratio for $\epsilon_2 = 10^{-3},10^{-2}$, and $10^{-1}$ from left to right, using Eq.(\ref{eqn:ZBR}) with $\alpha_d = \alpha_{\rm EM} = 1/127$.}
\label{fig:zcs}
\end{figure}

{\bf Tevatron and LHC:} The cross-section for production of a pair of dark sector states off the $Z^0$ at Tevatron and LHC is shown in Fig.~\ref{fig:zcs}. To a good approximation, the interfering diagram involving an off-shell dark vector-boson can be neglected since it produces final states which are too soft. With several ${\rm fb^{-1}}$ of data, the Tevatron can already probe this production channel for ${\rm BR}(Z^0\rightarrow {\rm dark~sector}) > 10^{-6}$. The LHC should be sensitive to the same parameter region with several hundreds ${\rm pb^{-1}}$ of data.  

\subsection{Electroweak-ino Pair Production}

Electroweak-ino pair production is well understood and has been discussed extensively in the literature.  In particular, the tri-lepton final state with its low SM background offers one of the better discovery channels for the conventional MSSM \cite{Baer:1992dc}. 

In order to minimize the hadronic activity it is useful to focus on the production of colorless particles. While strong production channels yield much larger cross-sections, they result in more complicated events.  At this early stage of the investigation, we prefer to concentrate on the lepton jets by
themselves. At the Tevatron and LHC, the colliding particles certainly carry color, so the possibilities are few. One possibility is the Drell-Yan production of sleptons. Another is the pair production of electroweak-inos, which are superpositions of winos, binos, and higgsinos. After these -inos are  produced, they will promptly cascade decay through the visible sector until they reach the lightest superpartner of the MSSM, which in turn decays into dark sector superpartners.  The precise details of the MSSM cascade depend on the exact MSSM spectrum\footnote{In this work, we will limit ourselves to no more than a single step by assuming that the spectrum contains at most one additional ino below the ones produced. It is important to realize that since DM is no longer associated with the MSSM per say, any of the MSSM sparticles can be the lightest as long as it is unstable and can decay into the dark sector.}, but we will assume that it only produces isolated leptons, as is the case for example in decays of the chargino to the neutralino. 

\begin{figure}[h]
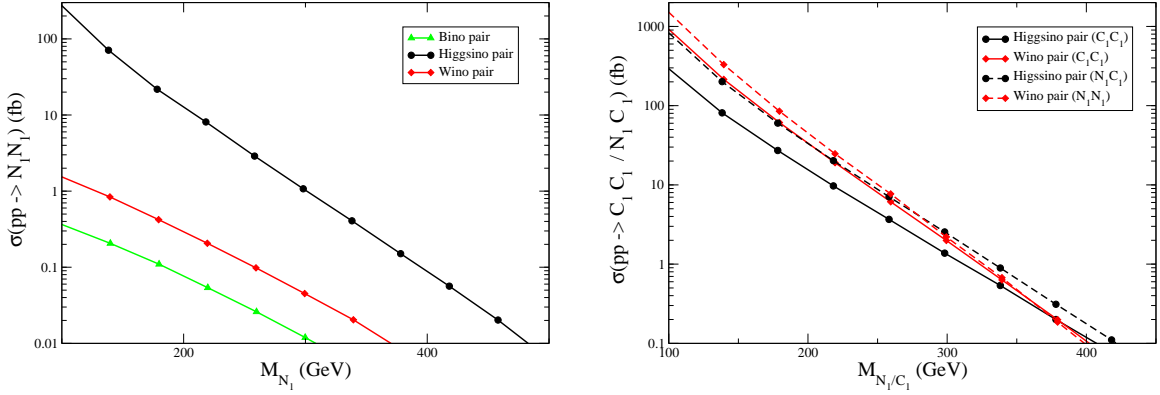

\begin{center}
\includegraphics[scale=.3]{neupair_TEV.eps}\quad\quad
\includegraphics[scale=.3]{neuchg_TEV.eps}
\end{center}
\caption{Production cross-sections for the different ino states at the Tevatron. The left pane includes neutralino pair production for the different gauginos. The right pane shows chargino pair production as well as neutralino-chargino associated production. A squark mass of $750\GeV$ was assumed. Cross-sections were computed with Pythia \cite{Sjostrand:2006za}.}
\label{fig:inoprodTEV}
\end{figure}
\begin{figure}[h]
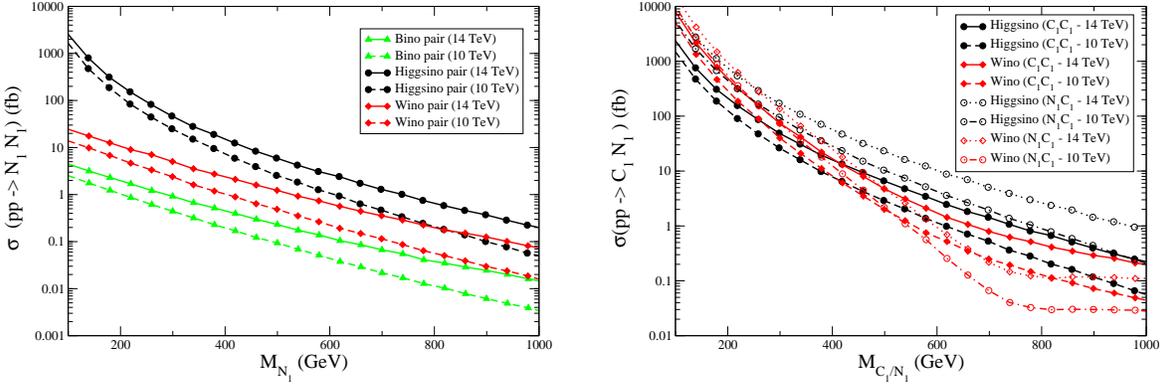

\begin{center}
\vspace{5mm}
\includegraphics[scale=.29]{neupair_LHC.eps}\quad\quad
\includegraphics[scale=.29]{neuchg_LHC.eps}
\end{center}
\caption{Same as Fig.~\ref{fig:inoprodTEV}, but for the LHC with center of mass energy of $14\TeV$ as well as $10\TeV$.}
\label{fig:inoprodLHC}
\end{figure}

We divide the electroweak production channels into three categories, neutralino-pair, chargino-pair, and neutralino-chargino associate production. In Figs.~\ref{fig:inoprodTEV} and \ref{fig:inoprodLHC} we depict the production cross-sections for the different pure gaugino states at the Tevaton and LHC, respectively. The purpose of these graphs is to give an estimate for the cross-sections involved for general MSSM spectra rather than concentrate on a particular benchmark scenario. In the case of neutralino pair production, the pure bino and wino states are produced only through a t-channel exchange of a squark, and their production cross-section is therefore suppressed compared with the higgsino state which enjoys its coupling to the $Z^0$.  Charged wino pair production has a much larger cross-section due to the large coupling of charged winos to the electroweak vector-bosons. For the purpose of computing the neutralino-chargino associate production we assume a degeneracy between the charged and neutral state, however, in realistic spectra those are usually split and the cross-sections are somewhat modified.

Both Tevatron and early LHC data should be sensitive to large parts of the parameter space. It is clear that generically, -ino production is dominated by channels with at least one chargino. This fact has important implications on the type of event topologies we can expect to encounter in association with lepton jets. As the chargino cascades down to the lightest neutralino it will emit a hard isolated lepton or other SM particles as we now discuss. 

\subsection{Neutralino Decays into the Dark Sector}
\label{sec:neutdecay}

Once produced, electroweak-inos will promptly cascade down to the lightest neutralino, $\tilde{N}_1$. In the process, they may emit leptons, quarks, or $Z^0/W^\pm$ (on-shell or off-shell) depending on the precise MSSM mass spectrum. Therefore, some of the events contain isolated leptons in addition to the lepton jets generated at a later stage of the event. Such isolated leptons should not cause any complications in a properly inclusive search. The cascades to $\tilde{N}_1$ may also result in colored particles and hence QCD-jets, but for the purpose of the current study we will assume that a substantial fraction of such cascades result in no colored particles. This is not a strong assumption as it is satisfied in many concrete examples of MSSM spectra \cite{Allanach:2002nj}, and may even be discarded altogether in actual lepton jet searches including QCD-jets. 

\begin{figure}[h]
\begin{center}
\includegraphics[scale=1]{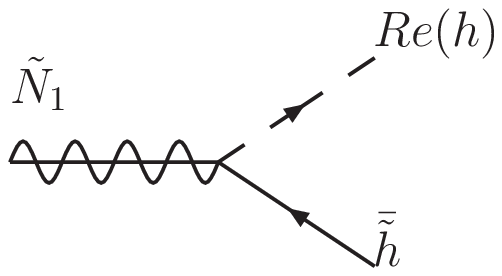}\quad\quad
\includegraphics[scale=1]{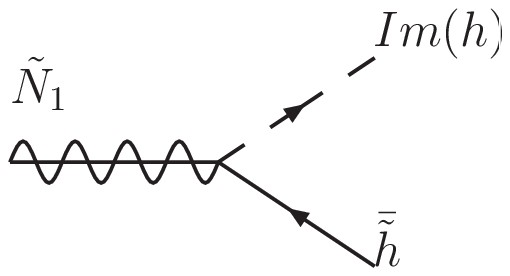}
\end{center}
\caption{The MSSM neutralino can decay to the light elements in the dark sector. We distinguish between the real and imaginary part of the dark higgses, because one linear combination is in fact the goldstone boson eaten by the broken dark gauge group. Thus, it can decay directly into lepton pairs while the other higgses cannot.}
\label{fig:neutdecay}
\end{figure}

The couplings in eq.~\ref{eq:portal} imply that the lightest neutralino decays into the dark sector via its bino or wino fraction.  The higgsino state is rendered unstable through its mixing with the wino and bino, as depicted in Fig.~\ref{fig:neutdecay}.  The complex scalar fields shown in this figure are dark gauge eigenstates separated into real and imaginary parts. The precise linear combinations of fields which define the mass eigenstates are described in the appendix. In particular, one combination is the eaten goldstone boson of the dark gauge symmetry.  As we will see, it is conceptually simpler to work in the gauge basis, especially when describing the ensuing radiation. 

The event shape is determined to a large degree by production, since the electroweak-inos are produced almost at threshold. Therefore, the lightest neutralino $\tilde{N}_1$ is \textbf{not} very boosted in the lab frame and so its decay products are \textbf{not} collimated. Each of the two neutralinos gives off a pair which yields a total of four highly-boosted dark sector particles which are well-separated in the lab frame. They form the seeds of the dark showers described in the next section. The final outgoing lepton jets will consist of some isolated leptons coming from the electroweak-ino cascades together with up to four lepton jets created by the boosted dark sector states.

We conclude this section by considering some special cases in which the decay into the dark sector proceeds slightly differently, resulting in alternative final configurations. For instance, in low-scale SUSY breaking scenarios, the gravitino is light and so the lightest neutralino can decay into a gravitino with the emission of a real photon.  This process competes with neutralino decays into the dark sector. Generally, those two decay rates are vastly different \cite{Baumgart:2009tn}, but in some instances they may be comparable \cite{ShihThomas}. If so, the events will contain a single hard photon, lepton jets and possibly other isolated leptons. This topology was the basis for a recent analysis looking for lepton jets in the Tevatron \cite{Abazov:2009hn}.

Second, there are cases where the degeneracy between the charged and neutral winos is lifted only by electromagnetic loop corrections, as is the case for instance in anomaly mediated supersymmetry breaking. The decay $\tilde{W}^\pm \rightarrow \tilde{W}^0 + \pi^\pm$  proceeds with an approximate lifetime of $10^{-10}~{\rm sec}$ and results in a displaced vertex. Thus, the lepton jets that emerge out of the decay of $\tilde{W}^0$ into the dark sector reconstruct this displaced vertex and are themselves displaced. We do not consider such scenarios further in this work, but only comment that lepton jets with displaced vertices may require special consideration when designing an experimental search because triggers usually assume the leptons to come from the primary interaction point\footnote{We thank Roger Moore for pointing this out to us.}.

\section{Dark Sector Dynamics}
\label{sec:dynamics}

In this section we discuss the evolution of dark sector states that arise from visible sector decays.  As observed in \cite{Baumgart:2009tn} these states can seed long cascades if the dark sector happens to have a rich particle spectrum.  Such a scenario is generic if the dark sector has non-abelian gauge dynamics.  Along and at the end of these cascades, numerous leptons are produced from the resulting dark gauge bosons, yielding highly collimated lepton jets.  A primary observation of this paper is that dark sectors with relatively sparse spectra, for example abelian theories, also provide a sizeable production of lepton jets due to soft particle emission.  

To see this, consider the scenario described in the previous section, where each neutralino arising from a supersymmetric cascade in the visible sector decays into a well-separated and boosted dark scalar and fermion pair.  Each of these carries an energy of roughly half the neutralino mass ($\sim100's\GeV$).   Furthermore, since these dark sector states are so highly boosted (and have mass $\lesssim\GeV$) they will radiate soft and collinear dark photons with high probability \cite{Meade:2009rb,Bai:2009it}.  Each dark photon brehmsstrahlung eventually decays into a pair of leptons, and contributes substantially to the final lepton multiplicity of the resulting lepton jet.  This part of the evolution is universal and depends only very weakly on the precise details of the symmetry breaking. 

Once the shower completes, we dress the dark scalars and fermions with the proper mass eigenstates as detailed below in section \ref{sec:cascades}. Depending on the final state, there may or may not be a pair of leptons at the termination point of the dark sector cascade.  For example, if the final state is the eaten goldstone boson of the dark symmetry breaking, then it will decay to a lepton pair directly.  If instead it is a scalar which lacks a direct coupling to leptons, then it decays through an intermediate (on-shell or off-shell) dark vector-boson (see Fig.~\ref{fig:decaymodes}). 

In this section we discuss each stage of the dark sector evolution:  showering, dressing, and their combined effect on the structure of the lepton jet.  For our analysis we employ the simple abelian model described in the appendix with the necessary field content to spontaneously break the dark gauge symmetry.  As such, much of the abelian model phenomenology is indicative of a broad class of models, including more complicated extensions such as non-abelian theories.

\subsection{Dark Showering}
\label{sec:darkshower}
Decays into the dark sector produce highly boosted dark particles which radiate soft dark photons.  The number of photons radiated off a dark higgs, $h$, or dark higgsino, $\tilde h$, can be understood parametrically in terms of the Sudakov double logarithm:
\begin{equation}
\label{eq:sudakov}
N_{b_\mu} \sim \frac{\alpha_d}{2 \pi} \log \left( \frac{M_{\rm EW}^2}{M_{\rm dark}^2}\right)^2 \simeq 1.4 \left( \frac{\alpha_d}{0.1} \right) \end{equation}
which is the expected number of soft dark photon emissions occurring within an energy window defined by $M_{\rm dark}$ and $M_{\rm EW}$.  More precisely, $M_{\rm EW}$ is the invariant mass of the initial dark sector state, for example the mass of the $Z^0$ or $\tilde N_1$ that decays into the dark sector, and $M_{\rm dark}$ is the the scale of dark sector masses, which regulates the soft and collinear divergences.  For the estimate in the final step of Eq.~\eqref{eq:sudakov}, we have taken $M_{\rm EW} / M_{\rm dark} \sim 10^2$.  For $M_{\rm dark} \sim$ GeV, the only parameters that determine the amount of showering are therefore $\alpha_d$ and $M_{\tilde N_1}$.  Since the ratio of scales, $M_{\rm EW} \gg M_{\rm dark}$, is independent of the details of the GeV-scale spectrum, showering is a universal process.  Furthermore, due to the logarithmic enhancement, it is clear that showering is an $\cO(1)$ effect, even for weakly coupled $\uoned$.  

In order to perform a detailed study of dark sector evolution, we have implemented the virtuality ordered parton shower employed by Sherpa \cite{Krauss:2005re} and Pythia \cite{Sjostrand:2006za}.  This algorithm is given by the repeated application of the following: 1) given an on-shell ``mother'' particle, sample the Sudakov form factor to obtain the mother's new virtuality $t$ and the energy fraction $z$ (and $1-z$) going to each of its daughters, and 2) re-shuffle kinematics to be consistent with four-momentum conservation.  We use the kinematic expressions of \cite{Bauer:2007ad}.  For a substantially more detailed description of the virtuality ordered parton shower, refer to the above references.

For a weakly coupled dark sector with a GeV-scale mass gap, we have found several simplifications to the parton shower which have little effect on lepton jet observables.  First, we note that the dark shower is dominated by splitting functions with both a soft and collinear divergence, $h \rightarrow h \gamma^\prime$ and $\tilde h \rightarrow \tilde h \gamma^\prime$.  Other splittings are only enhanced by the collinear log, and therefore constitute a $\lesssim 10\%$ effect.
Second, massive splitting functions and the precise GeV-scale virtuality cutoff have little effect.  This is because the amount of radiation depends only logarithmically on $M_{\rm dark}$.  We have found that varying the virtuality cutoff of the shower by an order of magnitude around a GeV only has a noticeable effect on the $p_T$ distribution of radiated dark photons for $p_T \lesssim 5$ GeV, where the dependence is $\cO(1)$.  Third, the running of the dark gauge coupling is negligible, since the theory is weakly coupled in the relatively small range between $M_{\rm EW}$ and $M_{\rm dark}$, and also constitutes a $\lesssim 10\%$ effect on the amount of showering.  Based on the above discussion, we adopt several simplifications for what follows, allowing us to keep the showers as model independent as possible.  In particular, we only include splitting functions that are double log enhanced, we neglect massive splitting functions and fix the virtuality cutoff to 1 GeV, and we neglect the running of the dark gauge coupling.  

Lastly, we mention a notable difference between abelian and non-abelian theories pertaining to angular ordering.  In particular, consider a soft emission of the form $A\rightarrow BC$.  In the case that $B$ and $C$ are both charged (for instance in a non-abelian theory), then any subsequent soft emission at large transverse wavelength from either $B$ or $C$ is suppressed by a Chudakov-like effect.  This is due to interference in the matrix element, and is simulated in virtuality ordered parton showers by enforcing `angular ordering,' where subsequent emissions are required to occur at smaller openings angles than prior emissions.  In our case, however, the theory is abelian, and the dominant process consists of a hard dark-charged $h$ or $\tilde h$ line emitting dark photons.  Since dark photons are neutral, there is no suppression arising from angular ordering.

\begin{figure}[h]
\begin{center}
\includegraphics[scale=.3]{ZradN.eps}
\includegraphics[width=0.44\textwidth, height=0.242\textheight]{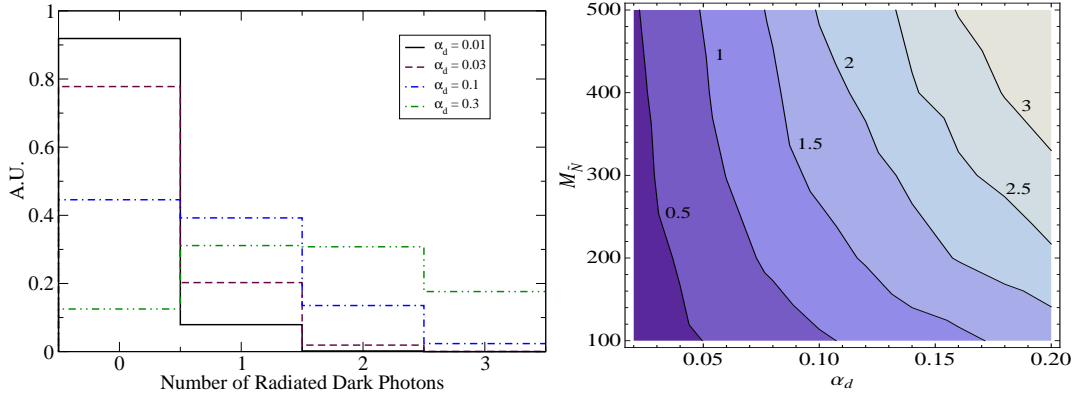}
\end{center}
\caption{On the left we depict the normalized distributions for the expected number of radiated dark photons in rare $Z^0$ decays into the dark sector. On the right is a contour plot of the number of soft dark photon emissions per neutralino as a function of the dark gauge coupling $\alpha_d$ and the neutralino mass $M_{\tilde N}$. The linear dependence on $\alpha_d$ and logarithmic dependence on $M_{\tilde N}$ is in accord with the naive expectation, Eq.~(\ref{eq:sudakov}). The plots were produced using a 3 GeV $p_T$ cut on the dark photons. Both plots are for LHC at 10 TeV center of mass energy.}
\label{fig:numofphotons}
\end{figure}

\begin{figure}[h]
\begin{center}
\includegraphics[scale=.35]{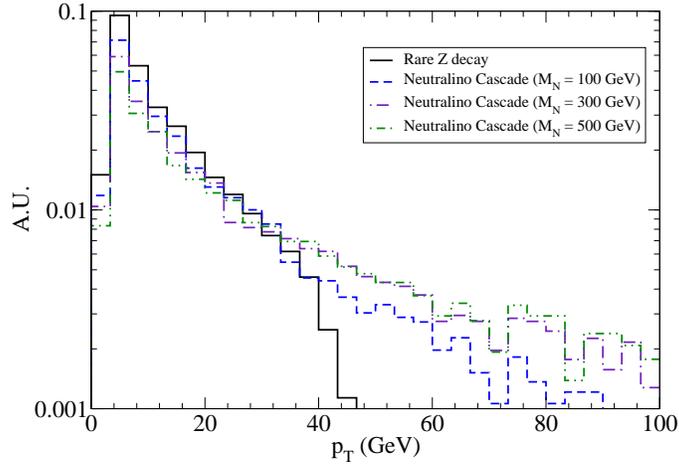}
\end{center}
\caption{The $p_T$ distribution of radiated dark photons for rare $Z^0$ decays and neutralino cascades. }
\label{fig:pTdist}
\end{figure}

In Fig.~\ref{fig:numofphotons} we present the average number of radiated dark photons in rare $Z^0$ decays and neutralino pair production events. The amount of radiation increases with the dark coupling as can be expected. It also increases with larger neutralino mass since the initial dark higgs and higgsinos are more energetic. In Fig.~\ref{fig:pTdist} we plot the $p_T$ distributions for the radiated photons for different parameters. 

\vspace{23mm}

Altogether, the effects of radiation in rare $Z^0$ decays into the dark sector are rather mild. The energy scale involved is somewhat lower as compared with neutralino pair production and subsequently the radiation is softer and less pronounced. On the other hand, for neutralino cascades into the dark sector the effects of radiation are important. The presence of these extra radiated photons modifies and enriches the structure of the resulting lepton jets as we discuss in the next section. 

\subsection{Decay of dark states back into leptons and pions}
\label{sec:cascades}

Once the virtuality reaches the dark state's mass ($\sim\GeV$) no further radiation is possible and the state is placed on-shell. At this point, it may decay back into light standard model particles such as leptons and pions\footnote{In what follows we mention leptons only for the sake of brevity, but pions are also possible. We discuss this issue more carefully below.} via the kinetic mixing operators, Eq. (\ref{eqn:kinmix}). The characteristics of the decay depend on the identity of the dark particle involved since not all are directly mixed with the standard model. For illustrative purposes we consider the two dark scalar model presented in the appendix, in which case the bosonic spectrum consists of a heavy dark higgs, $H_d$, a dark pseudo-scalar $a_d$, a dark vector-boson $b_\mu$ and a lighter dark higgs $h_d$. The fermionic spectrum might involve the superpartners of these particles, the dark higgsinos and gauginos. 

\begin{figure}[h]
\begin{center}
\includegraphics[scale=.7]{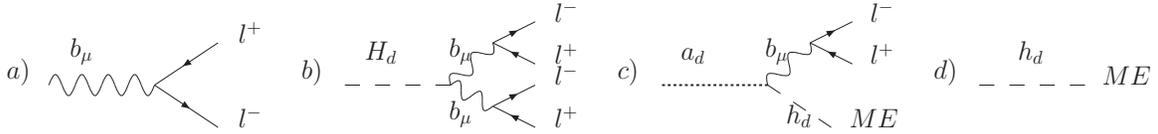}
\end{center}
\caption{The different decay modes associated with the dark bosons: a) $b_\mu$ direct decay into leptons through kinetic mixing; b) $H_d$ decays into two on-shell $b_\mu$ which then decay into two leptons each; c) $a_d$ decays into $h_d$ and dileptons through an on-shell or off-shell dark vector-boson depending on the detailed spectrum; d) $h_d$ if lighter than the other bosons typically decays outside the detector and constitutes missing energy.}
\label{fig:decaymodes}
\end{figure}

Since only $b_\mu$ mixes with hypercharge, it alone can decay directly into leptons. $H_d$ on the other hand we take to first decay into two on-shell $b_\mu$'s which later also decay, resulting in 4 leptons. $h_d$ can also decay through two $b_\mu$'s, but if it is lighter than the dark vector-boson then the decay is off-shell and results in a very long lifetime~\cite{Baumgart:2009tn, Batell:2009yf}. This typically means that $h_d$ escapes detection and is counted as missing energy. The pseudo-scalar, $a_d$, can typically decay into $h_d$ and two leptons through either on-shell or off-shell $b_\mu$. These different decay modes are shown in Fig.~\ref{fig:decaymodes}. While the specifics of the decay are fairly model dependent, we believe that the resulting phenomenology it describes is fairly universal in that each dark particle can do one of three things: 1) decay into an even multiplicity of leptons (2,4,6 . . . leptons); 2) constitute missing energy; 3) a combination of the two. 

In the next section we analyze the structure of lepton jets that results from the decay of the dark states back into leptons. The lepton jets are produced by first simulating the hard production process, followed by the decay into the dark sector, followed by a simulation of the dark radiation. Finally, in the last step we randomly assign the bosons to $H_d$, $b_\mu$, $a_d$, or $h_d$ and decay them accordingly, thus producing lepton jets. We take the dark fermions to be stable and only account for their radiation. 

\subsection{Lepton jets' morphology}

Before moving on to discuss the results of the simulations, we briefly discuss the qualitative differences between the resulting lepton jets and the type of jets one can expect from QCD. Lepton jets differs from QCD jets by both composition and shape.

The composition of the lepton jet is affected by the ratio of electrons/muons to pions which is determined by the decay modes of the dark vector-boson. The decay branching ratio of the vector-boson $b_\mu$ into pions is dictated by the electromagnetic form-factor at $q^2=m_b^2$, also known as the $R$ ratio~\cite{Amsler:2008zzb}. If the vector-boson mass is very close to the $\rho$-meson resonance, then its decay is mostly into pions and speaking of ``lepton jets'' is not very appropriate. In general, however, a sizable branching fraction into leptons can be expected and in what follows we consider several benchmarks with different branching fractions (${\rm Br}(b_\mu \rightarrow \pi^+\pi^-) = 1/7, 1/3, 3/5$ and taking the muon vs. electron branching fraction to be equal.).  This is an important effect to model because pion contamination will reduce the efficiency for lepton jet searches with hadronic isolation cuts as we discuss in the next section. 

Regarding the shape of lepton jet, they are usually made of a "hard core" of tightly packed high energy leptons coming from the primary dark bosons, and a "soft-shell" of somewhat sparse and less energetic lepton pairs coming from dark radiation. This configuration is quite different from usual QCD jets and can be used to search for this objects as we discuss in the next section. The "soft-shell" itself may contain additional discriminating power. Since the dark photon is relatively light $\lesssim\GeV$, even the soft radiation usually results in fairly collimated leptons. We therefore expect the pattern of energy deposition in the "soft-shell" to be of fairly isolated hits. 

It is important to realize that there are certain regions of parameter space where the resulting jets look very similar to QCD jets and their discovery is difficult at best. As we just mentioned, when the dark vector-boson mass is very close to the $\rho$-meson mass, it decays mostly into pions. If the dark coupling is very large, radiation is a substantial effect which will smear out the distinction between the "hard core" and "soft shell". In this case, the resulting jets resemble QCD jets in both composition (mostly mesons) and shape (smeared energy deposition). That said, in most other parts of parameter space, lepton jets are sufficiently different from QCD jets by both composition and shape that their discovery should be possible.

\section{Lepton Jets and Experimental Searches}

In this penultimate section we present the results of simulations for the different processes discussed above. Production of the dark states was simulated using Madgraph~\cite{Alwall:2007st}. The later cascade and decay back into the SM was simulated with a private code using Mathematica. 
\label{sec:lepjet}

\subsection{Lepton jets from rare $Z^0$ decays}

\begin{figure}[h]
\begin{center}
\includegraphics[scale=0.5]{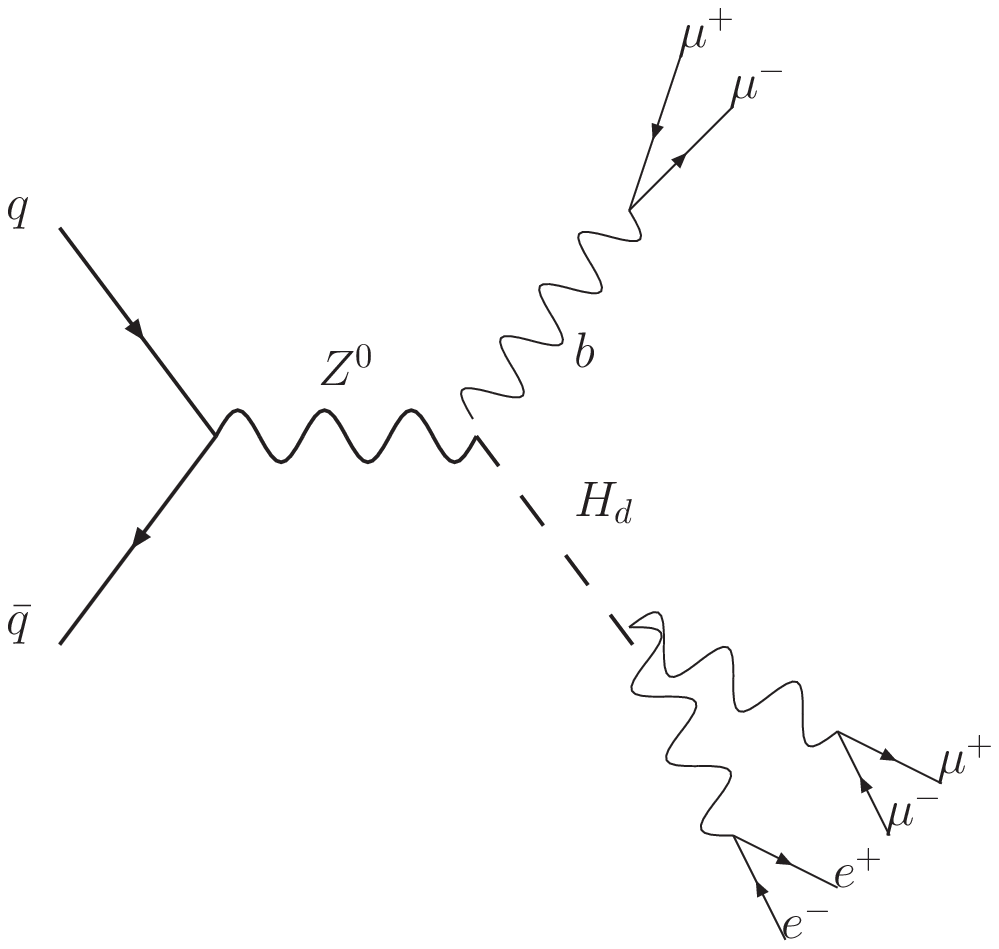}\hspace{2mm}
\includegraphics[scale=0.5]{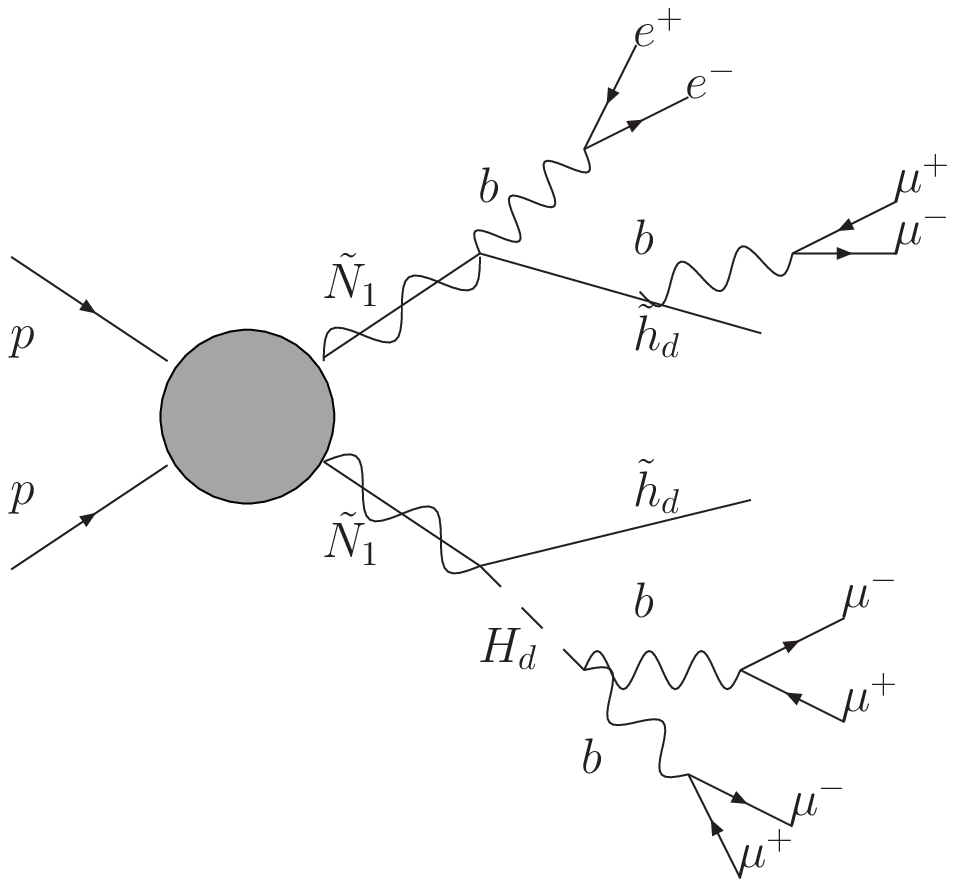}
\end{center}
\caption{The event topologies discussed in section \ref{sec:lepjet}. On the left we depict a rare $Z^0$ decay into dark states which subsequently decay as in Fig.~\ref{fig:decaymodes}. This would result in two isolated lepton jets recoiling against each other. On the right is a neutralino pair production event with the neutralino ultimately decaying into the dark sector. We allow the dark bosons to decay as usual, but keep the dark fermions stable, aside from possible radiation. The events therefore consist of 2 hard lepton jets, missing energy, and softer leptons coming from radiation of the dark fermions (radiation from the dark bosons would normally be clumped together with the harder leptons coming from the cascade.). }
\label{fig:ZandNeutEvents}
\end{figure}

As discussed in the previous section, the effects of dark radiation in the case of rare $Z^0$ decays into the dark sector are mild with possibly one radiated dark photon in a fraction of the events (left pane of Fig.~\ref{fig:numofphotons}). The structure of the lepton jets in this case is therefore straightforward to understand. If the decay is into dark fermions, the event contains large amounts of missing energy and possibly one or two dileptons from the fermions cascade. On the other hand the decay into dark bosons lead to 4 distinct topologies. Assuming CP conservation, the $Z^0$ must decay into one CP-even and CP-odd boson. On one side of the event we can expect either $b_\mu$ or $a_d$ and on the other side $H_d$ or $h_d$. Their decays are depicted in Fig.~\ref{fig:decaymodes} and the corresponding lepton jets are comprised of 2 or 4 collimated leptons with some of the events containing missing energy. Radiation will increase the lepton multiplicity in some of the events, but should not substantially affect the event topology and lepton jet structure. We consider this an important channel for lepton jet searches since it is rather clean with a simple event topology as depicted on the left of Fig.~\ref{fig:ZandNeutEvents}.  

\subsection{Lepton jets from neutralino cascades}

In the case of neutralino cascades, shown on the right of Fig.~\ref{fig:ZandNeutEvents}, the effects of dark radiation are more substantial. If not for radiation, one would expect two clean lepton jets in almost every event, coming from the scalars decaying into leptons (with the exception of $h_d$ which would constitute missing energy, and any decays involving pions). Including radiation, the dark fermions may also produce lepton jets (albeit softer) which would increase the lepton jet multiplicity. On the other hand, dark radiation might also deteriorate the signal. In particular, a justified concern is the pollution of lepton jets by pions coming from radiated dark photons. Such contaminants will make the distinction between lepton jets and QCD-jets difficult.

\begin{figure}[h]
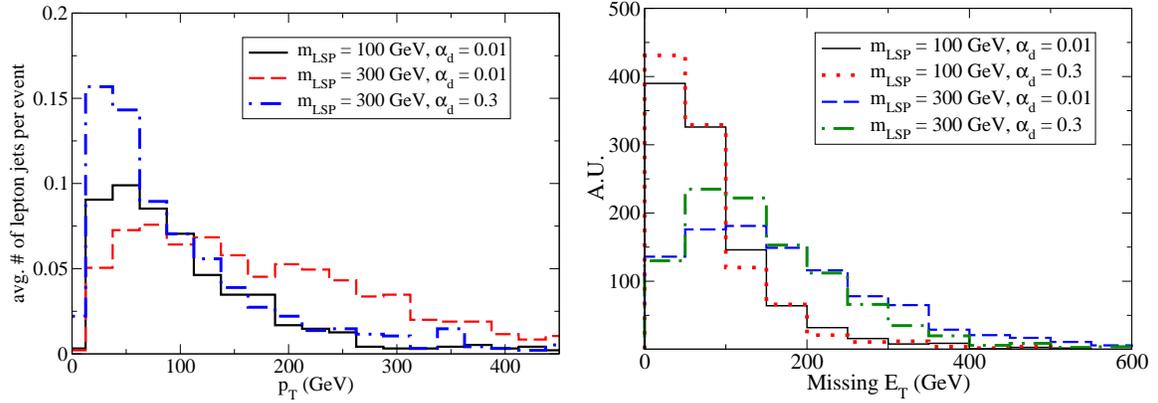

\begin{center}
\vspace{5mm}
\includegraphics[scale=0.3]{pT_LJ.eps}\hspace{2mm}
\includegraphics[scale=0.3]{etmiss.eps}
\end{center}
\caption{On the left pane we plot the differential lepton jet number per event as a function of their $p_T$ for different values of the neutralino mass and gauge coupling. Increasing the mass tends to increase the average $p_T$ as expected, while a larger gauge-coupling tends to soften it because more radiation is generated. On the right pane we show the missing energy distribution in the events. The distributions are fairly insensitive to the dark gauge coupling. }
\label{fig:LJplots}
\end{figure}

In order to investigate these effects in detail, we simulated the entire process, starting from production, going through the dark states' radiation and cascade, and ending with the decay back into leptons and pions. We assume that the neutralino always decays into the dark sector via its coupling to the dark current, Eq. (\ref{eq:portal}). In simulating the cascades we included all 4 decays shown in Fig.~\ref{fig:decaymodes} in equal proportions and kept the dark fermions stable. The results are presented in Tables \ref{tbl:numLepJet100} and \ref{tbl:numLepJet300}  where we present the probability of finding clean lepton jets per event for different values of the dark gauge-coupling and the branching ratio into pions. Clean lepton jets are defined as follows: at least two leptons with $p_T>10\GeV$ each in a cone of $\Delta R < 0.1$ with hadronic \textit{and} leptonic isolation of $\sum p_T < 3\GeV$ in $0.1< \Delta R < 0.4$. The number in brackets represents the same probability, but where leptonic isolation has been removed and only hadronic isolation is required. Not surprisingly, the efficiency without lepton isolation is higher compared with the case where isolation is required. This is especially so when $\alpha_d$ is large and more radiation is expected in the isolation annulus. This effect may be important when looking for light sectors which are strongly coupled and radiate copiously. On the left of Fig.~\ref{fig:LJplots} we show the differential lepton jet number per event as a function of the lepton jet $p_T$. As the dark coupling increases, the $p_T$ tends to decrease since the stronger radiation dilutes the hard core of the lepton jet.  On the right of Fig.~\ref{fig:LJplots} we depict the missing energy distribution in these events. It depends strongly on the mass of the neutralino and only weakly on the dark gauge coupling. 

\begin{table}[h]
\begin{center}
\begin{tabular}{|c||c|c|c||c|c|c||c|c|c||c|c|c||}
\hline
\multicolumn{7}{|c|}{Lepton Jet Efficiencies} \\
\hline
 & \multicolumn{3}{|c|}{1 Lepton-Jet} & \multicolumn{3}{|c|}{2  Lepton-Jet} \\
\hline
 \backslashbox{ \hspace{5mm}$\alpha_d$}{Br$_{b\rightarrow \pi\pi}$} &1/7 & 1/3 & 3/5 & 1/7 & 1/3 & 3/5 \\
 \hline
 0      & \text{0.46 (0.46)} & \text{0.36 (0.36)} & \text{0.26 (0.26)} & \text{0.18 (0.18)} & \text{0.08 (0.08)} & \text{0.02 (0.02)} \\
 0.01 & \text{0.46 (0.47)} & \text{0.39 (0.39)} & \text{0.26 (0.26)} & \text{0.15 (0.15)} & \text{0.1 (0.11)} & \text{0.03 (0.03)} \\
 0.03 &  \text{0.41 (0.42)} & \text{0.37 (0.37)} & \text{0.25 (0.26)} & \text{0.14 (0.17)} & \text{0.09 (0.1)} & \text{0.03 (0.03)} \\
 0.1   &  \text{0.39 (0.41)} & \text{0.36 (0.37)} & \text{0.21 (0.24)} & \text{0.14 (0.18)} & \text{0.06 (0.1)} & \text{0.02 (0.02)} \\
 0.3   & \text{0.31 (0.38)} & \text{0.27 (0.37)} & \text{0.17 (0.25)} & \text{0.07 (0.21)} & \text{0.04 (0.11)} & \text{0.02 (0.03)}\\
 \hline
\end{tabular}
\end{center}
\caption{Clean lepton jet efficiencies for different values of the dark gauge-coupling and ${\rm Br}(b\rightarrow\pi^+\pi^-)$. The neutralino mass was set to $\tilde{M} = 100$ GeV. For $\alpha_d=0$ dark radiation was switched off. The number of lepton jets increases with $\alpha_d$ as radiation becomes more likely. The requirement for ``clean'' lepton jets, as described in the text, results in a decrease in efficiency with the growth of the branching ratio into pions. In brackets are efficiencies for the case where only hadronic isolation is required in the $0.1 < \Delta R < 0.4$ annulus. The statistical error on the efficiencies is $\pm0.03$}
\label{tbl:numLepJet100}
\end{table}%

\begin{table}[h]
\begin{center}
\begin{tabular}{|c||c|c|c||c|c|c||c|c|c||c|c|c||}
\hline
\multicolumn{7}{|c|}{Lepton Jet Efficiencies} \\
\hline
 & \multicolumn{3}{|c|}{1  Lepton-Jet} & \multicolumn{3}{|c|}{2  Lepton-Jet}  \\
\hline
 \backslashbox{ \hspace{5mm}$\alpha_d$}{Br$_{b\rightarrow \pi\pi}$} &1/7 & 1/3 & 3/5 & 1/7 & 1/3 & 3/5  \\
\hline 
0      &  \text{0.49 (0.49)} & \text{0.47 (0.47)} & \text{0.31 (0.31)} & \text{0.28 (0.28)} & \text{0.14 (0.15)} & \text{0.05 (0.05)} \\
0.01 & \text{0.47 (0.47)} & \text{0.44 (0.45)} & \text{0.31 (0.32)} & \text{0.3 (0.31)} & \text{0.16 (0.16)} & \text{0.04 (0.04)} \\
0.03 &  \text{0.43 (0.41)} & \text{0.47 (0.48)} & \text{0.3 (0.3)} & \text{0.27 (0.3)} & \text{0.14 (0.16)} & \text{0.04 (0.05)} \\
0.1   & \text{0.43 (0.39)} & \text{0.41 (0.44)} & \text{0.29 (0.32)} & \text{0.23 (0.3)} & \text{0.13 (0.18)} & \text{0.05 (0.07)} \\
0.3   & \text{0.38 (0.32)} & \text{0.34 (0.36)} & \text{0.25 (0.34)} & \text{0.16 (0.3)} & \text{0.11 (0.22)} & \text{0.05 (0.09)} \\
\hline
\end{tabular}
\end{center}
\caption{Same as Table \ref{tbl:numLepJet100}, but with $\tilde{M} = 300$ GeV}
\label{tbl:numLepJet300}
\end{table}%

\begin{figure}[t]
\begin{center}
\vspace{5mm}
\includegraphics [width=0.44\textwidth, height=0.242\textheight]{numLJ_threshold.eps}
\includegraphics[width=0.44\textwidth, height=0.242\textheight]{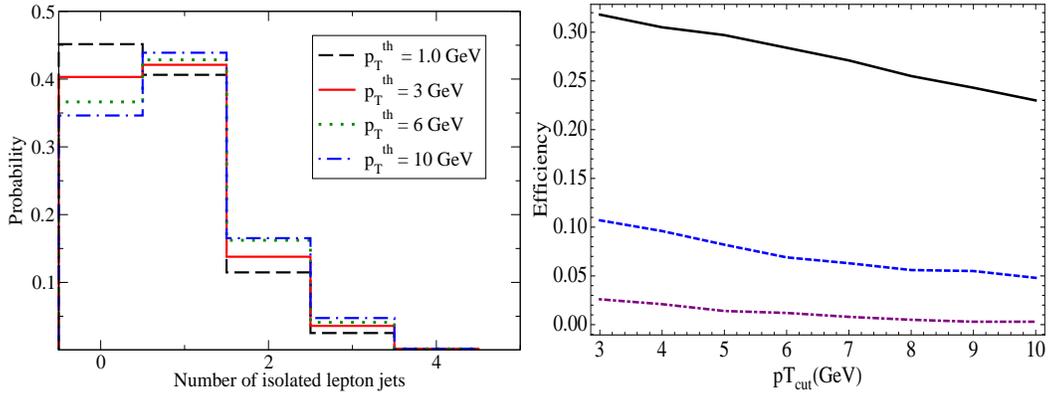}
\end{center}
\caption{On the left pane we show the dependence of the lepton jet efficiency on the isolation cut $\sum p_T$. On the right pane we plot the efficiency for 2 (black-solid), 3 (blue-dashed), and 4 (purple-dotted) clean lepton jets as a function of the $p_{T_{\rm cut}}$ of the second hardest lepton in the jet. The parameters used are $\tilde{M} = 300$ GeV, $\alpha_d = 0.1$, and ${\rm Br}(b\rightarrow\pi^+\pi^-)=1/7$. This plot shows that by lowering the $p_T$ requirement on the second lepton in the lepton jets, one can increase the efficiency for their observation. }
\label{fig:secondlepton}
\end{figure}

The probability of finding lepton jets decreases with increasing dark gauge-coupling since more radiation results in a lower average $p_T$ and softer leptons. A larger branching ratio into pions tends to pollute the lepton jets. At  ${\rm Br}(b\rightarrow\pi^+\pi^-) = 3/5$, for example, it becomes very unlikely to observe more than one lepton jet per event. The probability of observing more than two lepton jets is rather small throughout the parameter space and is not depicted. However, it is important to realize that this is to some extent a consequence of the strict definition of clean lepton jets. Such a strict definition is probably necessary to trigger and search inclusively for the harder lepton jets, but it may be desirable to relax  the requirements somewhat for the other lepton jets in the events. In Fig.~\ref{fig:secondlepton} we plot the probability of finding more than one lepton jet as a function of the $p_T$ cut on the second hardest lepton in the jet. 

\begin{figure}[h]
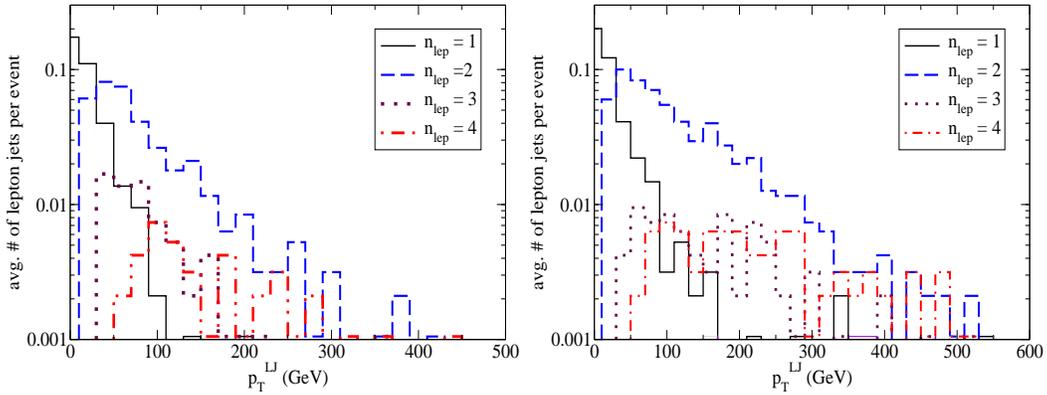

\begin{center}
\vspace{5mm}
\includegraphics [width=0.44\textwidth, height=0.242\textheight]{pTLJ-numLep-M1a4p2.eps}
\includegraphics[width=0.44\textwidth, height=0.242\textheight]{pTLJ-numLep-M3a4p2.eps}
\end{center}
\caption{The average number of lepton jets with a given number of leptons in them $n_{lep} = 1,2,3$, and $4$ for $M_{\tilde{N}_1} = 100\GeV$ ($300\GeV$) on the left (right) with $\alpha_d = 0.1$ and ${\rm Br}(b\rightarrow\pi^+\pi^-)=1/3$. Most lepton jets contain 2 leptons. The $n_{lep} = 1$ case does not really satisfy the requirements needed to be identified as a lepton jet, but is included here to illustrate that isolated leptons in those events are fairly soft. They originate from soft radiated dark photons which have decayed into two, fairly separated leptons. This does not include the possibility of a hard isolated lepton originating from chargino to neutralino cascade, a possibility which was not simulated here.}
\label{fig:LJnum}
\end{figure}

Another important characteristic of a lepton jet is its lepton multiplicity. Higher multiplicity helps with background rejection since few standard model processes can give more than 2 hard and collimated leptons. In Tables \ref{tbl:LepMulti100} and \ref{tbl:LepMulti300} we show the lepton multiplicity distribution in the hardest jet for different values of the gauge-coupling and branching fraction into pions. A sizable fraction of the events ($\sim \mathcal{O}(10\%)$) have at least one lepton jet with 4 hard leptons in it coming mostly from $H_d$ decay. The number of such lepton jets diminishes as the gauge-coupling increases because dark radiation pollutes the annulus $0.1 < \Delta R < 0.4$. This reduction is less significant if one removes the requirement of leptonic isolation in the $0.1 < \Delta R < 0.4$ region. In Fig. \ref{fig:LJnum} we show the differential distribution of lepton jets with a given number of leptons as a function of $p_T$.

\begin{table}[t]
\begin{center}
\begin{tabular}{|c||c|c|c||c|c|c||c|c|c||}
\hline
\multicolumn{10}{|c|}{Lepton Multiplicity in Clean Lepton Jets} \\
\hline
 & \multicolumn{3}{|c|}{2 Leptons} & \multicolumn{3}{|c|}{4 Leptons} & \multicolumn{3}{|c|}{6 Leptons} \\
\hline
 \backslashbox{ \hspace{5mm}$\alpha_d$}{Br$_{b\rightarrow \pi\pi}$} &1/7 & 1/3 & 3/5 & 1/7 & 1/3 & 3/5 &1/7 & 1/3 & 3/5 \\
\hline
0        & 0.41 & 0.31 & 0.21 & 0.22 & 0.13 & 0.06 & 0. & 0. & 0. \\
0.01  & 0.37 & 0.33 & 0.23 & 0.23 & 0.15 & 0.05 & 0. & 0. & 0. \\
0.03  &  0.33 & 0.31 & 0.22 & 0.2 & 0.14 & 0.06 & 0.01 & 0. & 0. \\
0.1    & 0.29 & 0.26 & 0.18 & 0.18 & 0.12 & 0.04 & 0.02 & 0.01 & 0. \\
0.3    & 0.22 & 0.19 & 0.14 & 0.09 & 0.08 & 0.03 & 0.02 & 0.01 & 0.\\
\hline
\end{tabular}
\end{center}
\caption{Lepton multiplicity in the hardest lepton jet for different values of the dark gauge-coupling and ${\rm Br}(b\rightarrow\pi^+\pi^-)$. The neutralino mass was set to $\tilde{M} = 100$ GeV. Odd number of leptons is fairly unlikely since the vector-boson always decays into a pair of leptons.}
\label{tbl:LepMulti100}
\end{table}%

\begin{table}[h]
\begin{center}
\begin{tabular}{|c||c|c|c||c|c|c||c|c|c||}
\hline
\multicolumn{10}{|c|}{Lepton Multiplicity in Clean Lepton Jets} \\
\hline
 & \multicolumn{3}{|c|}{2 Leptons} & \multicolumn{3}{|c|}{4 Leptons} & \multicolumn{3}{|c|}{6 Leptons} \\
\hline
 \backslashbox{ \hspace{5mm}$\alpha_d$}{Br$_{b\rightarrow \pi\pi}$}  &1/7 & 1/3 & 3/5 & 1/7 & 1/3 & 3/5 &1/7 & 1/3 & 3/5 \\
\hline
0        &  0.49 & 0.44 & 0.29 & 0.28 & 0.17 & 0.07 & 0. & 0. & 0. \\
0.01  & 0.53 & 0.43 & 0.29 & 0.25 & 0.18 & 0.06 & 0. & 0. & 0. \\
0.03  &  0.47 & 0.46 & 0.29 & 0.26 & 0.16 & 0.06 & 0.01 & 0.01 & 0. \\
0.1    & 0.42 & 0.43 & 0.32 & 0.25 & 0.16 & 0.06 & 0.04 & 0.02 & 0. \\
0.3    & 0.35 & 0.38 & 0.34 & 0.21 & 0.11 & 0.05 & 0.07 & 0.04 & 0.01\\
\hline
\end{tabular}
\end{center}
\caption{Same as Table \ref{tbl:LepMulti100}, but for $\tilde{M}=300$ GeV.}
\label{tbl:LepMulti300}
\end{table}%

\newpage

\subsection{Experimental Searches}

These findings suggest that it is possible to look for lepton jets in a fairly inclusive fashion and compel us to put forward some suggestions for such experimental searches. One such search was in fact already done in the Tevatron~\cite{Abazov:2009yi} where a fairly similar definition for lepton jets as the one discussed below was employed. We believe that many of the conclusions arrived at here will remain true and carry over in the case of a non-abelian dark gauge group as long as it does not condense. It is useful to try and arrive at some quantitative definition of a lepton jet and we suggest the following\footnote{We thank the participants of Boost09 and especially, B. Demirkoz, A. Haas, R. Moore,  
P. Schuster, N. Toro, and J. Wacker for very fruitful discussions concerning these issues.}: \\

\hspace{3mm}\parbox{140mm}{\textit{Two or more leptons each with $p_T>10\GeV$ inside a cone of $\Delta R < 0.1$ with hadronic/leptonic isolation cut of $\sum p_T < 3\GeV$ in an annulus of $0.1< \Delta R < 0.4$ around the lepton jet.}}\\	

We view this definition as a template to which more adequate figures can be added later once issues of backgrounds, triggering and other experimental difficulties are thoroughly analyzed and resolved. In particular, it will be important to bring the following parts under control:
\begin{enumerate}
\item ``Two or more leptons'' - Muons likely suffer from less background than electrons (either instrumental or physical) so we may require fewer of them than we do electrons. When only two hard leptons are present, one might demand the hardest two leptons to have the same flavor and opposite sign as this may prove useful in reconstructing the tracks as they separate in the magnetic field\footnote{We thank Andy Haas for pointing this out to us.}. This might be more difficult to do in the case of $H_d$ decay (or non-abelian scenarios) where more than one pair of hard leptons is expected.  	

\item ``$p_T > 10\GeV$'' - This figure can be changed depending on the lepton species under consideration. Also, it may be refined to allow for a softer secondary lepton. 

\item ``$\Delta R < 0.1$'' - This figure may be tuned depending on the process under consideration and the expected size of the opening angle. Also, one may want to define a lower limit in order to reduce background. 

\item ``$\sum p_T < 3\GeV$'' - A hadronic/leptonic isolation cut around the lepton jet core is necessary in order to fight background. It may be possible to relax the requirement for leptonic isolation if the background is not too large. The precise annulus in which this isolation cut is required (tentatively, $0.1 < \Delta R < 0.4$) should be thought of carefully especially since dark radiation leads to more dispersed lepton jets. 

\item ``around the lepton jet'' - In this work we defined the cone as follows: We first used an iterative algorithm to build the lepton jet 4-vector. Starting with the hardest lepton we collected all leptons within $\Delta R =0.1$ around it and added their 4-vectors to the lepton jet. This was repeated until no further leptons were found within $\Delta R =0.1$ around the lepton jet 4-vector. This same 4-vector was then used to define the isolation cone $0.1 < \Delta R < 0.4$. This choice was motivated by the physics since the combined 4-vectors reconstruct the dark state. However, other definitions are possible, for example the one employed by Ref.~\cite{Abazov:2009yi} where several isolation cones were defined, one about each lepton in the lepton jet. 
 
\item Lepton jets can have a displaced vertex as well. This case should be carefully studied by itself since in general the trigger algorithms assume that electrons and muons have a track in the inner detector and therefore displaced lepton jets may fail to trigger\footnote{We thank Roger Moore for pointing this out to us.}. 
\end{enumerate}  

This definition is likely to be revised and modified once collider effects are studied and a better understanding of the different backgrounds is developed. Once a proper operational definition of a lepton jet is formed, and maybe several interesting subclasses are identified, it is possible to conduct inclusive searches for these objects:

\begin{enumerate}
\item Lepton jet recoiling against a QCD-jet would be an inclusive search for a prompt dark photon production \cite{Baumgart:2009tn}.

\item Two lepton jets recoiling against each other and reconstructing the $Z^0$ would be an interesting signal of rare $Z^0$ decays into the dark sector and can be looked for at LEP, Tevatron, and LHC. 

\item Two (or more) lepton jets together with missing energy and possibly other isolated final states (e.g. a muon, an electron, and etc.) can be the result of electroweak-ino production and their eventual cascade into the dark sector.

\item Lepton jets in association with QCD-jets could be the result of strong production of colored particles which eventually cascade into the dark sector. This production mode was not investigated in this paper, but is important to investigate in future works on the subject since it enjoys a very large cross-section. 
\end{enumerate}

\section{Conclusions}

In this paper we investigated the (supersymmetric) production of dark states through electroweak processes. We carefully included the dark radiation that accompanies highly boosted dark states as well as their cascade evolution back into leptons and pions. This allows for a more realistic characterization of the resulting lepton jets and their properties. In particular, we put forward an operational definition for lepton jets that should aid in performing inclusive searches for these objects. We simulated the relevant processes (without including detector effects) and conclude that the efficiency for identifying clean lepton jets is rather high throughout the parameter space, even when possible pollution from pions and other leptons is included. 

The entire process depends on several disjoint assumptions and we have attempted to keep the discussion modular throughout. The production of dark states depends of course on the existence of the dark sector and its coupling to the standard model. The dark radiation is a result of the $\sim\GeV$ scale gauge group and depends on the gauge coupling and (logarithmically) on the energy carried by the dark states. The cascade back into the standard model depends on the details and spectrum of the dark sector which in this work we took to consist of two dark chiral multiplets (scalars and fermions). Finally, the decay branching fraction into pions vs. leptons is determined by the electromagnetic form-factor at the dark gauge-boson mass and we have modeled this effect by considering several pion branching fractions. 

However, while our analysis depends on numerous assumptions, we believe that the resulting phenomenology is rather robust and depends mostly on the existence of such a light sector with a small, but not negligible, interaction with the standard model. As such, the present work should aid future searches for lepton jets in high energy colliders irrespective of the precise details of the light sector. 

\label{sec:conclusions} 

\begin{acknowledgments}
We would like to thank Kyle Cranmer, Bilge Demirkoz, Andy Haas, and Roger Moore for extremely valuable comments and discussions regarding this work. I. Y. would like to thank Tel-Aviv university physics department for its hospitality. L.-T. W. is supported by the NSF under grant PHY-0756966 and the DOE under grant DE-FG02-90ER40542.  J.~T.~R. is supported by an NSF graduate fellowship. I.Y. is supported by the NSF under grant PHY-0756966 and the DOE under grant DE-FG02-90ER40542 as well as the James Arthur fellowship.
\end{acknowledgments}

\appendix
\renewcommand{\theequation}{A-\arabic{equation}}
\setcounter{equation}{0}

\section{A Simple Abelian Model}
\label{app:model}

Throughout this work, we have been assuming that the WIMP is coupled to some abelian gauge group $\uoned$ which is broken and becomes massive at $\sim\GeV$. In this appendix we briefly describe the dark higgs sector responsible for the spontaneous breaking of the gauge symmetry. Our notation follows that of Ref. \cite{Gunion:1989we} and the vertices can be read off the MSSM vertices by restriction of the electroweak higgs doublets to the neutral higgses' interactions with the $Z^0$ vector-boson. We do not discuss issues of naturalness or other model building concerns~\cite{Cheung:2009qd,Katz:2009qq,Morrissey:2009ur}, but simply use the model below as a framework to describe the resulting phenomenology. 

The dark higgs scalar potential, including soft terms is given by,
 \begin{equation}
 V(h_1,h_2) = m_1^2 |h_1|^2 + m_2^2 |h_2|^2 + m_{12}^2 h_1 h_2 + \frac{\gd^2}{2}\left( |h_1|^2 - |h_2|^2 \right)^2 + {\rm h.c.}
 \end{equation}
where the soft masses include possible contributions from $\mu$-like terms. We can express these in terms of the mass eigenstates, $h_d$ and $H_d$ which are the light and heavy real scalars, $a_d$ which is the pseudo-scalar, and $b_d$ which is the Goldstone eaten by the gauge field $b_\mu$, 
\begin{eqnarray}
h_1 &=& v_1 + \frac{1}{\sqrt{2}}\left(H_d\cos\alpha - h_d\sin\alpha + i a_d\sin\beta - i b_d \cos\beta  \right) \\
h_2 &=& v_2 + \frac{1}{\sqrt{2}}\left(H_d \sin\alpha + h_d\cos\alpha + i a_d\cos\beta - i b_d \sin\beta  \right) 
\end{eqnarray}
where $v_{1,2}$ are the vacuum expectation values (VEV) that spontaneously break $\uoned$, and their ratio determines $\beta$ through $\tan\beta = v_2/v_1$. The angle $\alpha$ is determined through the mass matrix of the real scalars. It is convenient to adopt $\tan\beta$, $m_{12}^2$, and $\mgam$ as independent parameters. We then have, 
\begin{eqnarray}
m_{a_d}^2 &=& m_{12}^2\left( \tan\beta + \cot\beta \right)\\
m_{H_d,h_d}^2 &=& \frac{1}{2}\left(m_{a_d}^2+\mgam^2 \pm \sqrt{\left(m_{a_d}^2 + \mgam^2 \right)^2 - 4 \mgam^2 m_{a_d}^2\cos^22\beta} \right) \\
\cos2\alpha &=& -\cos2\beta\left( \frac{m_{a_d}^2 - \mgam^2}{m_{H_d}^2 - m_{h_d}^2}\right)
\end{eqnarray} 

Most importantly, the following hierarchy is always maintained $m_{h_d} \leq m_{a_d},m_{b} \leq m_{H_d}$, which strongly constrains the type of decays involved. The relation between $m_{a_d}$ and $m_b$ is constrained by the requirement that $m_b$ decays into leptons and not into $h_d$ and $a_d$, so $a_d$ cannot be taken too light compared with $m_b$. Therefore, in large regions of parameter space, the decays in the dark higgs sector are as shown in Fig.~\ref{fig:decaymodes}. 

When simulating the cascades in the dark sector we assumed equal branching ratio of the MSSM neutralino into the different dark states, which is a very good approximation considering the large hierarchy of scales. We used phase-space distributions for the decay and chose $m_H=1.1\GeV$, $m_b=0.5\GeV$ and $m_a=1\GeV$ for concreteness, however, the kinematics is determined mostly by the large energy scale involved, namely, the neutralino mass. 

\bibliographystyle{JHEP}
\bibliography{portal}
\end{document}